\def\BorderBox#1#2{\vbox{
                               #2%
			}}%
\def\InsertFigure[#1 #2 #3 #4]#5#6#7{%
	\epsfxsize=#6%
	\epsfysize=#7%
	\epsfbox[#1 #2 #3 #4]{#5}%
	}
\documentstyle[a4,epsf,12pt]{article}
\newcommand{\beq}{\begin{equation}}
\newcommand{\eeqn}{\end{eqnarray}}
\newcommand{\beqn}{\begin{eqnarray}}
\newcommand{\eeq}{\end{equation}}
\newcommand{\np}{Nucl.\ Phys. \underline}
\newcommand{\pl}{Phys.\ Lett. \underline}
\newcommand{\pr}{Phys.\ Rev. \underline}
\newcommand{\prl}{Phys.\ Rev.\ Lett.\underline}
\newcommand{\fbstat}{f_B^{\rm stat}}
\newcommand{\fbsstat}{f_{B_{\rm s}}^{\rm stat}}
\newcommand{\calp}{{\cal P}}
\newcommand{\kcrit}{\mbox{$\kappa_{\rm crit}$}}
\newcommand{\kstrange}{\mbox{$\kappa_s$}}
\newcommand{\mev}{{\rm MeV}}
\newcommand{\gev}{{\rm GeV}}

\newcommand{\dl}{\stackrel{\leftarrow}{D}}
\newcommand{\dr}{\stackrel{\rightarrow}{D}}

\newcommand{\plus}{\makebox[15pt][c]{$+$}}
\newcommand{\minus}{\makebox[15pt][c]{$-$}}
\newcommand{\figurebox}[2]{\fbox{\vbox to#2in{\hbox to #1in{\hfil} \vfil}}}
\newcommand{\errr}[2]{\raisebox{0.08em}{\scriptsize {$\;\begin{array}{@{}l@{}}
                          \plus\makebox[1.35em][r]{#1} \\[-0.12em] 
                          \minus\makebox[1.35em][r]{#2} 
                        \end{array}$}}}
\newcommand{\err}[2]{\raisebox{0.08em}{\scriptsize {$\;\begin{array}{@{}l@{}}
                          \plus\makebox[0.95em][r]{#1} \\[-0.12em] 
                          \minus\makebox[0.95em][r]{#2} 
                        \end{array}$}}}
\newcommand{\er}[2]{\raisebox{0.08em}{\scriptsize {$\;\begin{array}{@{}l@{}}
                          \plus\makebox[0.55em][r]{#1} \\[-0.12em] 
                          \minus\makebox[0.55em][r]{#2} 
                        \end{array}$}}}

\newcounter{myfigure}

\newlength{\captionwidth}
\begin{document}

\begin{titlepage}

\begin{flushright}
Edinburgh Preprint: 93/526\\
Southampton Preprint SHEP 92/93-24
\end{flushright}

\vspace*{5mm}

\begin{center}
{\Huge Quenched Heavy-Light Decay Constants}\\[15mm]
{\large\it UKQCD Collaboration}\\[3mm]

{\bf R.M.~Baxter, S.P.~Booth, K.C.~Bowler, S.~Collins, D.S.~Henty,
R.D.~Kenway, D.G.~Richards, H.P.~Shanahan, J.N.~Simone, A.D.~Simpson,
B.E.~Wilkes}\\ Department of Physics, The University of Edinburgh,
Edinburgh EH9~3JZ, Scotland

{\bf A.K.~Ewing, L.~Lellouch, C.T.~Sachrajda, H.~Wittig}\\
Physics Department, The University, Southampton SO9~5NH, UK

\end{center}
\vspace{5mm}
\begin{abstract}

We present results for heavy-light decay constants, using both
propagating quarks and the static approximation, in $O(a)$-improved,
quenched lattice QCD. At $\beta=6.2$ on a $24^3\times48$ lattice we
find $f_D=185\er{4}{3}\,({\rm stat})\,\err{42}{7}\,({\rm
syst})\,\,\mev$, $f_B=160\er{6}{6}\;\err{53}{19}\;\mev$,
$f_{D_s}/f_D=1.18\er{2}{2}$ and $f_{B_s}/f_B=1.22\er{4}{3}$, in good
agreement with earlier studies.  From the static theory we obtain
$\fbstat=253\err{16}{15}\;\errr{105}{14}\;\mev$.  We also
present results from a simulation at $\beta=6.0$ on a $16^3\times48$
lattice, which are consistent with those at $\beta=6.2$.  In order to
study the effects of improvement, we present a direct comparison of
the results using both the Wilson and the improved action at
$\beta=6.0$.

\end{abstract}

\end{titlepage}

\section{Introduction}
\label{sec:intro}
The leptonic decay constants, $f_P$, of pseudoscalar mesons composed
of a heavy and a light quark play an important r\^ ole in
weak-interaction phenomenology.  In particular $f_B$, or more strictly
$f_B\sqrt{B_B}$ (where $B_B$, the $B$ parameter of $B^0$--$\bar B^0$
mixing, is expected to be close to one), is one of the principal
unknown quantities needed for the determination of the $CP$-violating
phase in the Standard Model, as well as other properties of weak
decays. Lattice QCD offers the opportunity for a non-perturbative
computation of the operator matrix elements which are necessary for
the determination of the decay constants and $B$ parameters.

During the last few years there have been several lattice computations
of the decay constants of ``heavy-light" pseudoscalar (and vector)
mesons.  The results for the decay constant of the $D$ meson, obtained
using the Wilson action for the quarks, are in the region of
$200\;\mev$ (using a normalisation for which $f_\pi=$132 \mev). For
example, in his 1989 review S.\,Sharpe quoted \cite{sharpecapri}
$f_D\simeq 180\pm 25({\rm stat})\pm 30({\rm syst})\ \mev$ as his
summary of the lattice results.  More recent simulations with Wilson
fermions also give results in this range
\cite{allton2}--\cite{labrenz}.  The experimental bound is
$f_D<290\,\mev$ \cite{fdexperiment}.

In the heavy-quark limit the scaling law for the decay constant of a
heavy-light pseudoscalar meson is $f_P\sqrt{M_P}\sim$ constant (up to
mild logarithmic corrections). Lattice simulations using heavy-quark
masses in the charm region indicate that there are large corrections
to this scaling law (of order 40\% at the charm quark mass, decreasing
to about 15\% at the mass of the bottom quark)
\cite{allton2,abada,bernardfd}.  The value of the decay constant of
the $B$ meson deduced from these simulations is in the region of 180
\mev.  The conclusion that there are violations of the scaling law is
supported by the large value for $f_P\sqrt{M_P}$ deduced from
simulations obtained using a static (i.e infinitely-massive) heavy
quark~\cite{labrenz,eichten89,allton,wuppfbstat,apesmearing,duncan,alexandrou2}.

\par The important results and conclusions quoted above were obtained from 
simulations in which the mass of the heavy quark is large in lattice
units (up to about a half). One may therefore worry that
discretisation errors significantly contaminate the results. In this
paper we present the results for decay constants of heavy-light mesons
computed using the $O(a)$-improved lattice action proposed by
Sheikholeslami and Wohlert~\cite{sw}, with which the discretisation
errors in operator matrix elements (and hence in the computed values
of the decay constants) can be reduced from $O(m_Qa)$ to
$O(\alpha_sm_Qa)$, where $m_Q$ is the mass of the heavy quark
\cite{hmprs}. This formal reduction in discretisation errors
provides an important check on the stability of results and
conclusions obtained with Wilson fermions.
\par 
The results presented in this paper were obtained from two simulations
of quenched QCD, using the Sheikholeslami--Wohlert (SW) or ``clover''
fermion action for the quarks (see Subsection~\ref{subsec:improvement}
below).  Our main results come from a simulation on a $24^3\times 48$
lattice at $\beta = 6.2$, for which 60 gauge field configurations were
generated.  Details of this simulation and the determination of the
values of the Wilson hopping parameter corresponding to the chiral
limit, $\kcrit$, and to the mass of the strange quark have been presented
in ref.~\cite{strange}.  The heavy quarks have masses in the region of
the charm quark mass and we study the behaviour of the decay constants
with the mass of the heavy quark.  Interpolating to the mass of the
charm quark itself, and extrapolating the results to the mass of the
$b$ quark, we find that our best results for the decay constants of
the $B$ and $D$ mesons are
\beqn
f_D & = & 185\er{4}{3}\,({\rm stat})\,\err{42}{7}\,({\rm syst})\,\,\mev  
\label{eq:bestfd}\\ 
f_B & = & 160\er{6}{6}\,\err{53}{19}\,\,\mev  
\label{eq:bestfb}\\
\frac{f_{D_s}}{f_D} & = & 1.18\er{2}{2}\\
\frac{f_{B_s}}{f_B} & = & 1.22\er{4}{3}.
\eeqn
The details of this calculation and a complete set of results are 
presented in Section \ref{sec:fdprop}.
\par
The second simulation is on a $16^3\times 48$ lattice at $\beta=6.0$,
using 36 configurations. The results, which are consistent with those
mentioned above, are presented in Section \ref{sec:cmclover}.
In order to study the effects of improvement on the calculation of heavy-light
decay constants, 
we have repeated the computation for  
both the Wilson and SW actions using a subset of 16 of these
configurations.
The results and a discussion are presented in Subsection \ref{subsec:cmcomp}.
\par We have also computed $f_B$ in the static approximation (in
which contributions of $O(1/m_b)$ are neglected).
A discussion of the calculation and of the results is presented in 
Section \ref{sec:static}. The result from the simulation at $\beta=6.2$,
on 20 of the 60 configurations, is
\beq
\fbstat = 253\err{16}{15}\,({\rm stat})\,\errr{105}{14}\,({\rm syst})\ \mev,
\label{eq:bestfbstat}\eeq
and the result at $\beta=6.0$ on all 36 configurations is
\beq
\fbstat = 286\err{~8}{10}\,\err{67}{42}\,\,\mev.
\label{eq:bestfbstatsixo}\eeq

Finally, Section \ref{sec:concs} contains
our conclusions.

\subsection{Improved Action and Operators}
\label{subsec:improvement}
The SW action is
\begin{equation}
S_F^{SW}  = S_F^W - i\frac{\kappa}{2}\sum_{x,\mu,\nu}\bar{q}(x)
         F_{\mu\nu}(x)\sigma_{\mu\nu}q(x),
\label{eq:sclover}\end{equation}
where $S_F^W$ is the Wilson action:
\beq
S_F^W = \sum _x \Biggl\{\bar{q}(x)q(x)
             -\kappa\sum _\mu\Bigl[
            \bar{q}(x)(1- \gamma _\mu )U_\mu (x) q(x+\hat\mu )
+\ \bar{q}(x+\hat\mu )(1 + \gamma _\mu )
   U^\dagger _\mu(x)
            q(x)\Bigr]\Biggr\}.
\label{eq:sfw}\eeq
The decay constants of heavy-light pseudoscalar and vector mesons are
computed using lattice axial and vector currents as interpolating
operators. In order to obtain $O(a)$-improved matrix elements we use
``rotated'' operators \cite{hmprs}:
\beq
\bar Q(x)(1+\frac{1}{2}\gamma\cdot\dl)\,\Gamma\,
(1-\frac{1}{2}\gamma\cdot\dr)\,q(x),
\label{eq:impop}\eeq
where $\Gamma$ is one of the Dirac matrices (either $\gamma^\mu\gamma^5$ 
or $\gamma^\mu$), and $Q$ and $q$ represent the fields of the heavy and light
quark respectively.
\par In the static effective theory, in which the heavy propagator is 
expressed in terms of the link variables \cite{eichten}, it is
sufficient to rotate the light-quark fields only~\cite{borrelli2},
i.e. to use the operators
\beq
\bar Q(x)\,\Gamma\,(1-\frac{1}{2}\gamma\cdot\dr)\,q(x),
\label{eq:static_rotations}
\eeq 
in order to eliminate the $O(a)$-discretisation errors.


\subsection{Renormalisation Constants $Z_V$ and $Z_A$}
\label{subsec:zvza}
In order to determine the physical values of the decay constants from
those obtained in lattice simulations using the interpolating
operators in eq. (\ref{eq:impop}), it is necessary to know the corresponding
renormalisation constants. These are defined by requiring that
$Z_A A^{\rm latt}_\mu$
and $Z_V V^{\rm latt}_\mu$ are the correctly normalised currents, where the
superscript ``${\rm latt}$'' denotes that the operator is a lattice
current.
These renormalisation constants have been calculated at one-loop order in
perturbation theory for the SW action with rotated operators~\cite{borrelli}:
\beqn
Z_V & = & 1 - 0.10 g^2 \label{eq:zv}\\ Z_A & = & 1 - 0.02
g^2.\label{eq:za}\eeqn In this paper they are evaluated using the
``boosted'' coupling suggested in ref. \cite{lm}; specifically, we use
$g^2=6/(\beta\,u_0^4)$, where $u_0$ is a measure of the average link
variable, for which we take $u_0=(8\kcrit)^{-1}$.  It has been
suggested~\cite{lm,mackenzie} that the use of such an effective
coupling, rather than the bare lattice coupling, resums some of the
large higher-order corrections, and in particular some of the tadpole
diagrams.  Using the measured values of $\kcrit$ from our simulations,
we obtain $Z_A\simeq 0.97$ (0.96) and $Z_V\simeq 0.83$ (0.82) for the
simulation at $\beta$ = 6.2 (6.0).
\par In a recent non-perturbative determination of 
these renormalisation constants, obtained by requiring that the
correctly-normalised currents obey the continuum Ward Identities, it
was found that $Z_V=0.824(2)$ and $Z_A=1.09(3)$ \cite{mpsv}.  These
results were obtained from a simulation at $\beta=6.0$ for one value
of the quark mass.  It remains to be checked that the results are
independent of the quark mass and insensitive to small variations in
$\beta$.  For this reason, we use the perturbative values, given above,
throughout the paper.  We note, however, that the non-perturbative
value of $Z_A$ may be larger by about $15\%$.  In ref.~\cite{strange}
we obtained $f_\pi/m_\rho=0.138\er{6}{9}$, using the perturbative
value of $Z_A$.  A larger value of $Z_A$, such as $Z_A=1.09$, would
bring this result closer to the physical value of 0.172.  However, we
also observed that $f_K/f_\pi$, which does not require
$Z_A$, was in very good agreement with the experimental value, and
therefore we quote values for the ratios $f_D/f_\pi$ and $f_B/f_\pi$
in the following sections.
\par The normalisation of the axial current in the static effective theory
is discussed in Section \ref{sec:static}.

\subsection{Error Estimation}
\label{subsec:errors}

Statistical errors are obtained from a bootstrap procedure~\cite{EFRON}.
This involves the creation of 1000 bootstrap samples from the original
set of $N$ configurations by randomly selecting $N$ configurations per
sample (with replacement). Correlators are fitted for each bootstrap 
sample by minimising~$\chi^2$. During the fits, correlations among
different timeslices are taken into account, whereas correlations 
among different values of the quark mass are neglected. 
The latter correlations are preserved by using the same sequence
of bootstrap samples at each quark mass.
When extrapolating
our results to the chiral limit and physical meson masses, the correlation
matrix for the fitted quantities is estimated from the full
bootstrap ensemble.
All quoted statistical errors
are obtained from the central 68\% of the corresponding
bootstrap distributions~\cite{Oaimp}.

\par We attempt to quantify the systematic error 
arising from the uncertainty in the value of the lattice spacing, $a$,
determined from properties of light hadrons, and from the string
tension~\cite{strange}.  The differences between results obtained
using our central value for $a^{-1}\,[\gev]$ and our upper and lower
estimates are quoted as systematic uncertainties in the final
estimates for decay constants in physical units. Hereafter, where we
quote two errors, the first is statistical and the second is
systematic.


\subsection{Extended Interpolating Operators}
\label{subsec:smearing}
In order to isolate the ground state in correlation functions
effectively, it is useful to use extended (or ``smeared'')
interpolating operators for the mesons.  In particular, in the static
theory it has been found to be essential to use smeared operators in
order to obtain any signal for the ground state~\cite{boucaud}.  In
this study we use gauge-invariant Jacobi smearing on the heavy-quark
field (described in detail in ref.~\cite{smearing}), in which the
smeared field, $Q^S(\vec x,t)$, is defined by
\beq
Q^S(\vec x,t)\equiv \sum_{\vec x^\prime}J(x,x^\prime)Q(\vec x^\prime,t),
\label{eq:kernel}\eeq
where 
\beq
J(x,x^\prime)=\sum_{n=0}^N\,\kappa_S^n\Delta ^n(x, x^\prime)
\eeq
and
\beq
\Delta(x,x^\prime)=\sum_{i=1}^{3}\{\delta_{\vec x^\prime, \vec x-\hat\imath}
U^\dagger_i(\vec x-\hat\imath,t) + \delta_{\vec x^\prime, \vec x+\hat\imath}
U_i(\vec x,t)\}.
\eeq
Wuppertal smearing \cite{wuppsmearing}, which uses
the operator $(1 - \kappa_S \Delta)^{-1}$
as the kernel of the smearing,
corresponds to $N=\infty$, provided that $\kappa_S$ is sufficiently small
to guarantee convergence.
Following the discussion in ref.~\cite{smearing}, we choose $\kappa_S=0.25$
and use the parameter $N$ to control the smearing radius, defined by
\beq
r^2\equiv \frac{\sum_{\vec x}|\vec x|^2 |J(x, 0)|^2}
{\sum_{\vec x} |J(x, 0)|^2}.
\eeq
The values of $N$ and $r$ used in each of the calculations below will be
quoted in the corresponding sections.


\section{Decay Constants from the Simulation at $\beta =6.2$}
\label{sec:fdprop}

In this section we present the results obtained for the decay
constants of heavy-light mesons from our simulation on 60
configurations of a $24^3\times 48$ at $\beta=6.2$, using the SW
action in the quenched approximation.  The computations are performed
for four different values of the mass of the heavy quark,
corresponding to $\kappa_h=0.121$, 0.125, 0.129 and 0.133, and for
three values of the mass of the light quark, corresponding to
$\kappa_l=0.14144$, 0.14226 and 0.14262.  The mass of the charm quark
corresponds approximately to $\kappa_h=0.129$.  The value of the
hopping parameter corresponding to the mass of the strange quark is
$\kstrange$ = 0.1419\er{1}{1} and the critical value is
$\kcrit=0.14315\er{2}{2}$ \cite{strange}.

\par The decay constants are determined by computing two-point correlation
functions of the form 
\beq
C^{QR}_{J_1J_2}(t)=\sum_{\vec x}\langle0|J_1^Q(x)J_2^{\dagger\, R}(0)|0\rangle
\label{2ptfn}\eeq
where $J_1$ and $J_2^\dagger$ are interpolating operators which can
annihilate or create the pseudoscalar or vector meson being studied.
The labels $Q,R$ denote whether a local~($L$) or smeared~($S$)
interpolating operator is being used.  In this simulation we use
Jacobi smearing with $N=75$, corresponding to a smearing radius of
$r=5.2$.  The decay constants are obtained from the matrix elements of
the local operators, which are determined by computing both the
$C^{SS}$ and $C^{LS}$ correlation functions.

\par In order to determine the decay constant, it is necessary to know the
value of the lattice spacing in physical units. This can be done by
relating the lattice measurements of some dimensionful quantity to its
physical value, e.g. the mass of a light hadron or $f_\pi$. Among the
other frequently-used choices are the string tension,$\sqrt{K}$, and
the $1P-1S$ mass splitting in charmonium. Using $m_\rho$ to set the
scale in our study of light hadrons~\cite{strange} we found
$a^{-1}(m_\rho)=2.7(1)\,
\gev$, and a mass spectrum in physical units which was close
to experimental values. Furthermore, our determination of the string
tension
\cite{Oaimp} gave $a^{-1}=2.73(5)\,\gev$. Encouraged by the consistency of
these results, we use
\beq
a^{-1} = 2.7\,\gev.
\label{eq:scale}\eeq
\par However, the study described in ref.~\cite{strange} showed that the measurement
of the pion decay constant gave a higher value for the scale, i.e.
$a^{-1}(f_\pi)=3.4\er{2}{1}\,\gev$, using the perturbative value for
$Z_A$.  In order to get an estimate of the systematic uncertainties in
the final numbers, we evaluate all our results using the central value
of $a^{-1}(f_\pi)$ as well, and quote the difference as the upper
systematic error on decay constants. The lower systematic error is
obtained from the uncertainty of $-0.1\,\gev$ in $a^{-1}(m_\rho)$.

In an attempt to reduce the systematic errors associated with the
value of the renormalisation constant of the axial current, we also
compute $f_{D,B}/f_\pi$, and determine $f_{D,B}$ by using the physical
value of $f_\pi$.


\subsection{Decay Constants of Pseudoscalar Mesons} \label{psmesons}

In order to determine the pseudoscalar decay constants, we start by
fitting the two-point correlation function
\beqn
C^{SS}_{PP}(t)&\equiv&\sum_{\vec x}
\langle 0|P^S(\vec x,t)P^{\dagger S}(0)|0\rangle  
\nonumber\\ 
 & \rightarrow & \frac{Z_{P^S}^2}{2M_P}\exp(-M_PL_t/2)
\cosh\left( M_P(L_t/2-t)\right),
\label{eq:psps}\eeqn
where $P$ is the pseudoscalar density,
$Z_{P^S}=\langle0|P^S(0)|P\rangle$ and $L_t$ is the temporal extent of
the lattice.  This correlation function gives the best determination
of the masses of the heavy-light pseudoscalars.  Symmetrizing in
Euclidean time, the fitting range was chosen to be $13\leq t\leq 22$
for all three values of the light-quark mass.  Good plateaus in the
effective mass were observed and stable fits obtained.  The values of
the masses of the pseudoscalar and vector mesons for the twelve
$\kappa_h$-$\kappa_l$ combinations are presented in Table
\ref{tab:md}. The values obtained by linear extrapolation to the
chiral limit for the light quark are also tabulated.

\begin{table}
\centering
\begin{tabular} {|c|c|c|c|}\hline
$\kappa_h$ & $\kappa_l$ & $M_P$ & $M_V$ \\ \hline
       & 0.14144     & 0.924\er{3}{1} & 0.944\er{4}{2} \\ 
0.121  & 0.14226     & 0.900\er{3}{2} & 0.920\er{4}{3} \\ 
       & 0.14262     & 0.890\er{4}{3} & 0.909\er{5}{4} \\ 
       & $\kcrit$    & 0.875\er{4}{3} & 0.894\er{6}{4} \\ 
\hline
       & 0.14144     & 0.822\er{3}{1} & 0.847\er{4}{2} \\ 
0.125  & 0.14226     & 0.799\er{3}{2} & 0.823\er{4}{3} \\ 
       & 0.14262     & 0.789\er{4}{2} & 0.811\er{5}{4} \\ 
       & $\kcrit$    & 0.773\er{5}{2} & 0.797\er{6}{4} \\ 
\hline
       & 0.14144     & 0.715\er{3}{1} & 0.745\er{4}{2} \\ 
0.129  & 0.14226     & 0.691\er{3}{2} & 0.721\er{4}{3} \\ 
       & 0.14262     & 0.681\er{4}{2} & 0.711\er{5}{4} \\ 
       & $\kcrit$    & 0.665\er{5}{2} & 0.695\er{6}{4} \\ 
\hline
       & 0.14144     & 0.599\er{3}{1} & 0.637\er{3}{2} \\ 
0.133  & 0.14226     & 0.574\er{3}{2} & 0.613\er{4}{3} \\ 
       & 0.14262     & 0.564\er{4}{2} & 0.603\er{5}{4} \\ 
       & $\kcrit$    & 0.546\er{4}{2} & 0.588\er{6}{4} \\ 
\hline
\end{tabular}
\begin{center}
\begin{minipage}[t]{\captionwidth}
\caption{Masses (in lattice units) of the pseudoscalar and vector mesons 
for the twelve $\kappa_h$-$\kappa_l$ combinations at $\beta=6.2$
on a $24^3\times 48$ lattice.
Also presented are the values obtained by linear extrapolation to the chiral
limit ($\kappa_l\rightarrow \kcrit= 0.14315$).\label{tab:md}}
\end{minipage}
\end{center}
\end{table}

\par
\begin{figure}[t] 
\begin{center}
\leavevmode
\BorderBox{2pt}{%
\InsertFigure[140 480 480 800]{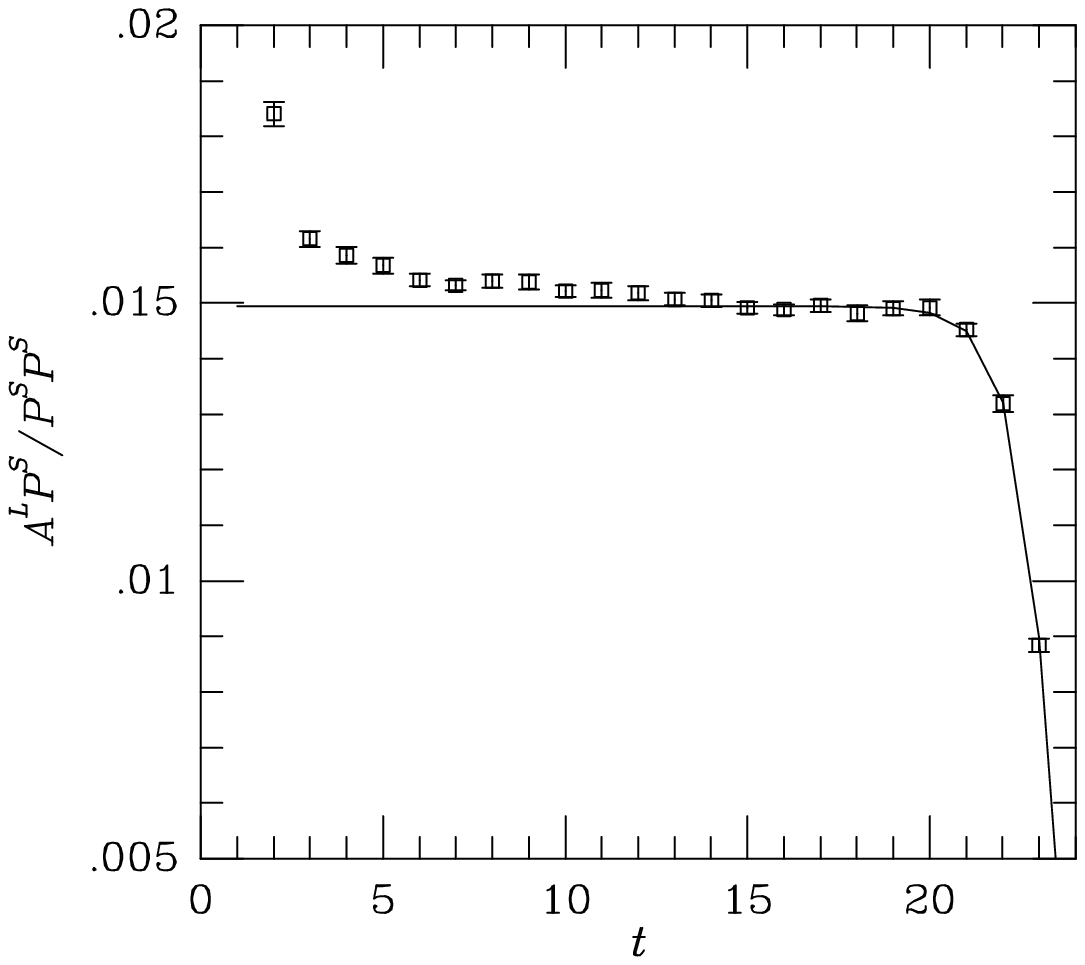}{340bp}{320bp}%
}
\begin{minipage}[t]{\captionwidth}
\caption{The ratio of correlators defined in eq.~(\protect\ref{eq:apoverpp})
plotted versus~$t$ for $\kappa_h=0.129$, $\kappa_l=0.14262$.  The
curve represents the fit using timeslices 15--22.\label{fig:apoverpp}
}
\end{minipage}
\end{center}
\end{figure}

 At large values of $t$, the ratio of correlation functions
\beq 
\frac{C_{AP}^{LS}(t)}{C_{PP}^{SS}(t)}\rightarrow
\frac{\langle0|A_4^L(0)|P\rangle}{\langle0|P^S(0)|P\rangle}
\tanh \left( M_P(L_t/2-t)\right)
\label{eq:apoverpp}\eeq
is used to extract the pseudoscalar decay constant, where $A_4$ is the
temporal component of the axial current. The ratio is fitted in the
range $15\leq t\leq22$ with the pseudoscalar mass $M_P$ (in each
bootstrap sample) constrained to its value extracted from fits to
eq.~(\ref{eq:psps}).  In Fig.~\ref{fig:apoverpp} we plot the ratio of
correlators together with the fit to eq.~(\ref{eq:apoverpp}) as a
function of~$t$. Using the value of $Z_{P^S}$ obtained from the fits
of eq.~(\ref{eq:psps}), the matrix elements of the local axial current
are obtained. Although there are other ways of determining these
matrix elements, we find that the ratios in eq.~(\ref{eq:apoverpp})
give the most precise determination.

%
%

\par In Table \ref{tab:fda} we present the results for the decay constants
(in lattice units) of the pseudoscalar mesons for the twelve 
$\kappa_h$-$\kappa_l$ 
combinations, as well as the values obtained by linearly extrapolating the 
results to the chiral limit.
We also tabulate the results for the quantity
$f_P\sqrt{M_P}$ which, in the heavy-quark limit, is independent of the mass
of the heavy quark (except for a mild logarithmic dependence). 

\begin{table}
\centering
\begin{tabular} {|c|c|c|c|c|}\hline
$\kappa_h$ & $\kappa_l$ & $f_P/Z_A$ & $f_P\sqrt{M_P}/Z_A$ 
           & $1/(f_V\,Z_V)$ \\ \hline
       & 0.14144    & 0.086\er{1}{1} & 0.083\er{2}{1} 
                    & 0.124\er{2}{3} \\ 
0.121  & 0.14226    & 0.079\er{1}{1} & 0.075\er{2}{1} 
                    & 0.116\er{3}{3} \\ 
       & 0.14262    & 0.076\er{2}{2} & 0.071\er{2}{2} 
                    & 0.111\er{3}{3} \\ 
       & $\kcrit$   & 0.071\er{2}{1} & 0.066\er{2}{1} 
                    & 0.105\er{4}{4} \\ 
\hline
       & 0.14144    & 0.086\er{1}{1} & 0.078\er{1}{1} 
                    & 0.141\er{3}{3} \\ 
0.125  & 0.14226    & 0.079\er{1}{1} & 0.070\er{1}{1} 
                    & 0.132\er{3}{3} \\ 
       & 0.14262    & 0.076\er{2}{2} & 0.067\er{2}{2} 
                    & 0.128\er{4}{4} \\ 
       & $\kcrit$   & 0.071\er{2}{1} & 0.062\er{2}{1}
                    & 0.123\er{4}{4} \\ 
\hline
       & 0.14144    & 0.085\er{1}{1} & 0.071\er{1}{1}
                    & 0.163\er{3}{3} \\ 
0.129  & 0.14226    & 0.078\er{1}{1} & 0.065\er{1}{1}
                    & 0.155\er{3}{3} \\ 
       & 0.14262    & 0.075\er{2}{1} & 0.062\er{1}{1}
                    & 0.151\er{4}{4} \\ 
       & $\kcrit$   & 0.071\er{2}{1} & 0.057\er{2}{1}
                    & 0.146\er{5}{5} \\ 
\hline
       & 0.14144    & 0.082\er{1}{1} & 0.063\er{1}{1} 
                    & 0.193\er{3}{4} \\ 
0.133  & 0.14226    & 0.076\er{1}{1} & 0.057\er{1}{1}
                    & 0.186\er{3}{4} \\ 
       & 0.14262    & 0.073\er{1}{1} & 0.055\er{1}{1}
                    & 0.183\er{5}{5} \\ 
       & $\kcrit$   & 0.069\er{1}{1} & 0.051\er{1}{1} 
                    & 0.179\er{5}{5} \\ \hline
\end{tabular}
\begin{center}
\begin{minipage}[t]{\captionwidth}
\caption{The decay constants (in lattice units) of the pseudoscalar 
and vector mesons. Also shown are the results for the combination 
$f_P\protect\sqrt{M_P}$ which in the heavy-quark limit is independent of the
heavy-quark mass (up to mild logarithmic corrections).\label{tab:fda}}
\end{minipage}
\end{center}
\end{table}

\begin{figure}[t]
\begin{center}
\leavevmode
\BorderBox{2pt}{%
\InsertFigure[140 480 480 800]{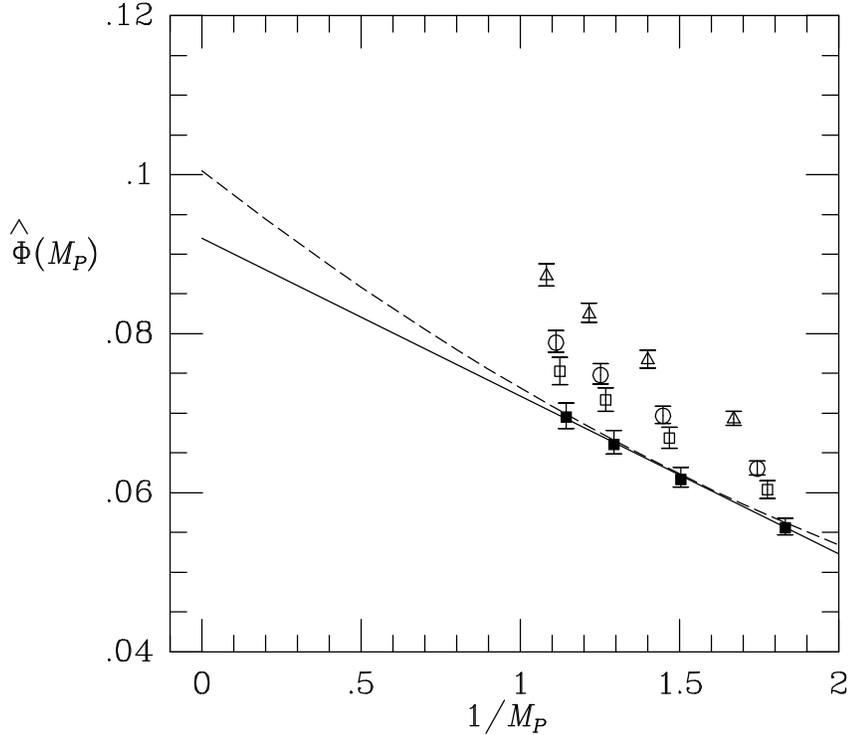}{340bp}{320bp}%
}
\begin{minipage}[t]{\captionwidth}
\caption{The data for $\protect\hat{\Phi}(M_P)$
plotted against the inverse meson mass. The open symbols denote points
with $\kappa_l<\kcrit$, whereas full symbols denote those extrapolated
to $\kcrit$. The solid line represents the linear fit to the
chirally-extrapolated points using the three heaviest meson masses,
whereas the dashed curve results from a quadratic fit to all
four.\label{fig:fdsqrtma}}
\end{minipage}
\end{center}
\end{figure}

We start the discussion of our results with the behaviour of the 
pseudoscalar decay constants as a function of the mass of the meson, 
with all dimensionful quantities given in lattice units. In the heavy-quark
limit, the quantity $f_P\sqrt{M_P}$ scales like
\beq
f_P\sqrt{M_P} = {\rm const.}\times \left[\alpha_s(M_P)\right]^{-2/\beta_0},
\quad M_P \longrightarrow\infty.
\label{eq:fP_scaling}
\eeq
In order to detect possible deviations from this scaling law we plot
in Fig.~\ref{fig:fdsqrtma} the quantity\footnote{The normalization
factor $\alpha_s(M_B)^{-2/\beta_0}$ is convenient when comparing these
results with those obtained in the static theory.}
\beq
\hat{\Phi}(M_P) \equiv (\alpha_s(M_P)/\alpha_s(M_B))^{2/\beta_0}Z_A^{-1}
f_P\sqrt{M_P}
\label{eq:phihat}
\eeq
as a function of $1/M_P$.  We approximate $\alpha_s(M)$ by
\beq
    \alpha_s(M)=\frac{2 \pi}{\beta_0 \log(M/\Lambda_{{\rm QCD}})}
\label{eq:alphas}
\eeq
where we take $\Lambda_{{\rm QCD}}=200\, \mev$, and $\beta_0 = 11 -
\frac{2}{3} n_f$, with $n_f = 0$ in the quenched approximation.
From the figure we see that $\hat{\Phi}(M_P)$
increases as the mass of the heavy 
quark is increased (in agreement with the behaviour found using the Wilson 
action for the quarks \cite{allton2,abada,bernardfd}). In order to quantify 
this behaviour, we fit $\hat{\Phi}(M_P)$ to either a linear or
quadratic function of
$1/M_P$:
\beq
\hat{\Phi}(M_P)=A\left( 1-\frac{B}{M_P}\right)
\label{eq:linfit}
\eeq 
or
\beq
\hat{\Phi}(M_P)=C\left( 1-\frac{D}{M_P}+\frac{E}{M_P^2}\right).
\label{eq:quadfit}
\eeq 
We have performed these fits twice; once using the values of $f_P\sqrt{M_P}$
for all four values of $\kappa_h$, and once using those for only the smallest
three $\kappa_h$'s (i.e. for the heaviest three heavy-quark masses).
The results of the fits are given in Table \ref{tab:lqfit}.
We find that the 
non-scaling corrections are of $O(30\%)$ for $f_D$ and $O(10\%)$ for 
$f_B$, in agreement with previous results obtained using Wilson fermions 
\cite{allton2,abada,bernardfd}.  From the quadratic fit to the data at all four 
heavy-meson masses we find, in physical units,
\beqn
Ca^{-3/2} & = & 0.45\err{~2}{~2}\err{19}{~3}\, \, \, \gev^{3/2}
\nonumber \\
Da^{-1} & = & 0.84\err{11}{~8}\err{22}{~3} \, \, \, \gev \nonumber \\
Ea^{-2} & = & 0.28\err{~7}{~9}\err{16}{~2} \, \, \, \gev^2.
\label{eq:physical_fit_params62}
\eeqn
The second error in eq.~(\ref{eq:physical_fit_params62}) corresponds
solely to the uncertainty in the scale. It should be mentioned that
ignoring the residual logarithmic dependence of $f_P\sqrt{M_P}$
on~$M_P$ makes the slope more pronounced.  However it is clear from
Fig.~\ref{fig:fdsqrtma} and Table~\ref{tab:lqfit} that the logarithmic
corrections to the scaling law can by no means account for the
observed slope in $f_P\sqrt{M_P}$.

\begin{table}
\centering
\begin{tabular}{|c||c|c||c|c|c|} \hline
 & \multicolumn{2}{c||} {Linear Fit} & \multicolumn{3}{c|}{Quadratic Fit} 
\\ \hline
  & $A$ & $B$ & $C$ & $D$ & $E$  \\ \hline
4 $\kappa_h$'s & 0.089\er{3}{2} & 0.199\err{5}{7}  
               & 0.101\er{5}{5} & 0.31\err{4}{3} & 0.038\err{9}{13}  \\
3 $\kappa_h$'s & 0.092\er{3}{3} & 0.216\err{9}{10}  
               & 0.102\er{6}{8} & 0.33\err{10}{5} & 0.045\err{18}{39}\\
\hline
\end{tabular}
\begin{center}
\begin{minipage}[t]{\captionwidth}
\caption{Values of the parameters of the linear and quadratic fits to the
behaviour of the pseudoscalar decay constants with the mass of the mesons
(as defined in the text).
\label{tab:lqfit}
}
\end{minipage}
\end{center}
\end{table}

%
%

\par We use the parameters of the fits in Table \ref{tab:lqfit} to make 
our predictions for the values of the decay constants $f_D$ and $f_B$. The
results corresponding to the four fits are presented in Table~\ref{tab:fdfb}.
\begin{table}
\centering
\begin{tabular}{|c||c|c||c|c|} \hline
 & \multicolumn{2}{c||} {Linear Fit} & \multicolumn{2}{c|}{Quadratic Fit} 
\\ \hline
  & $f_D$ & $f_B$ & $f_D$ & $f_B$ \\ \hline
4 $\kappa_h$'s  & 185\er{4}{3}\err{45}{7} & 149\er{5}{3}\err{52}{7} 
& 185\er{4}{3}\err{42}{7} & 160\er{6}{6}\err{53}{8}   \\ 
3 $\kappa_h$'s  & 186\er{4}{3}\err{41}{7} & 154\er{5}{4}\err{53}{8} 
& 185\er{4}{3}\err{42}{7} & 160\er{7}{7}\err{54}{7}   \\ 
\hline
\end{tabular}
\begin{center}
\begin{minipage}[t]{\captionwidth}
\caption{Values of the decay constants $f_B$ and $f_D$ in \mev, 
corresponding to the linear and quadratic fits.
\label{tab:fdfb}
}
\end{minipage}
\end{center}
\end{table}
From this table it is clear that there is a further systematic
uncertainty in~$f_B$ of about 11\,\,\mev\ from extrapolating using
either linear or quadratic fits.  In contrast to this, since we
interpolate to $m_D$, the results for~$f_D$ are very stable. It should
be emphasised that choosing a different value for $\Lambda_{{\rm
QCD}}$ (e.g. $\Lambda_{{\rm QCD}}=250\,\,\mev$), or for the anomalous
dimension (e.g., by taking $n_f = 4$), changes the results by only
about~$1\,\,\mev$.

Taking the results from the quadratic fit using all four
$\kappa_h$ values we find: 
\beqn f_D & = &
185\er{4}{3}\err{42}{7}\,\,\mev \label{eq:fdres}\\ 
f_B & = &
160\er{6}{6}\err{53}{19}\,\,\mev \label{eq:fbres}
\eeqn 
where we
have included the uncertainty of $11\;\mev$ from the extrapolations in
the systematic error quoted for $f_B$.  We take the results presented in
equations (\ref{eq:fdres}) and (\ref{eq:fbres}) as our best
estimates of the decay constants of the $D$ and $B$ mesons. 

\par In ref.~\cite{abada} it was found useful to use the pion decay constant,
$f_\pi$, to set the scale in the computations of the decay constants
of heavy-light mesons. By calculating $f_D/f_\pi$ and $f_B/f_\pi$ it
may be expected that some of the systematic errors cancel, since, in
particular, the ratios are independent of $Z_A$.  Our results for the
decay constants obtained in this way are, as expected, close to the
upper systematic error margins in Table~\ref{tab:fdfb}.  We find
\beqn
\Big(\frac{f_D}{f_\pi}\Big)\times132\,\,\mev &=& 232\err{12}{5}\,\,\mev 
\label{eq:fdfpi}\\
\Big(\frac{f_B}{f_\pi}\Big)\times132\,\,\mev &=& 201\err{12}{8}\,\,\mev.
\label{eq:fbfpi}
\eeqn

\par Finally in this subsection, we present our results for $f_{D_s}$ and
$f_{B_s}$. These are obtained by interpolating the measured values of the
decay constants given in Table \ref{tab:fda} to 
$\kappa_l=\kstrange= 0.1419\er{1}{1}$~\cite{strange}. The extrapolations in 
the heavy-quark masses are done as above. We find
\beqn
\frac{f_{D_s}}{f_D}& = & 1.18\er{2}{2} \label{eq:fdsfd}\\ 
\frac{f_{B_s}}{f_B}& = & 1.22\er{4}{3} \label{eq:fbsfb}.
\eeqn
In physical units we obtain
\beqn
f_{D_s} & = & 212 \er{4}{4}\err{46}{7}\,\,\mev\label{eq:fds}\\ 
f_{B_s} & = & 194 \er{6}{5}\err{62}{9}\,\,\mev\label{eq:fbs}.
\eeqn

Recently the first measurement of $f_{D_s}$ has been made by the WA75
collaboration \cite{wa75}, who found $f_{D_s}=(232\pm\, 45\pm\, 20
\pm\, 48\, )\mev$. Our result is in good agreement with the measured
value, and also with previous lattice calculations using Wilson fermions
\cite{abada,labrenz}.

\subsection{Decay Constants of Vector Mesons}
In this subsection we present our results for the decay constants of 
heavy-light vector mesons. These are defined by
\beq
\langle0|V_\mu|V\rangle\equiv\epsilon_\mu\frac{M_V^2}{f_V}
= Z_V\langle0|V_\mu^L(0)|V\rangle,
\label{eq:fvdef}\eeq
where $|V\rangle$ represents a state containing a vector meson $V$,
with mass $M_V$, polarisation vector $\epsilon_\mu$ and decay constant
$f_V$.  $V_\mu^L$ denotes the local lattice vector current, defined in
eq.~(\ref{eq:impop}), with $\Gamma = \gamma_\mu$, which has to be
multiplied by the renormalisation constant~$Z_V$. The vector
mass~$M_V$ is extracted from fits to the correlator
\beqn
C_{VV}^{SS} & \equiv & \sum_{j=1}^3\sum_{\vec{x}}\,
  \langle0|V_j^S(\vec{x},t)V_j^S(0)|0\rangle    \nonumber\\
& \rightarrow & - \frac{3 Z_{V^S}^2}{2M_V}\exp(-M_V\,L_t/2)
                \cosh(M_V(L_t/2-t)),
\eeqn
where $V_j^S$ is the $j$th spatial component of the smeared vector
operator and $Z_{V^S}$ is defined through
\beqn
\langle0|V_j^S(0)|V\rangle = \epsilon_j Z_{V^S}.
\eeqn
Fitting timeslices $14\leq t\leq23$, symmetrized, we obtain the vector
meson masses shown in Table~\ref{tab:md}.  In order to extract the
matrix element of the local vector current we fit the ratio
\beq
\frac{C_{VV}^{LS}(t)}{C_{VV}^{SS}(t)} \longrightarrow
-\frac{\sum_{j=1}^3\,\langle0|V_j^L(0)|V\rangle
\epsilon^{\ast}_j}{3 Z_{V^S}}
\eeq
to a constant in the fitting interval $15\leq t\leq23$.

The results for $1/f_V$ for the twelve $\kappa_h-\kappa_l$
combinations are presented in Table \ref{tab:fda}, together with those
obtained after extrapolation to the chiral limit.  The
chirally-extrapolated values for $f_V^{-1}\,Z_V^{-1}$ are now
interpolated to the $D^*$ mass using a quadratic fit to the data at
all four values of $\kappa_h$, giving,
\beq 
      f^{-1}_{D^*} = 0.110\er{5}{5}\err{36}{5}.
\eeq
This is slightly below, but still compatible with, earlier studies
(e.g.
\cite{abada}) when the systematic error is taken into account. This result
remains unaltered if a linear fit is used instead of a quadratic one.


\subsection{A Test of the Heavy Quark Symmetry}
In the heavy-quark limit,
the decay constants of heavy-light pseudoscalar and vector mesons are
related by~\cite{UMcorrections}
\beq
U(M)\equiv\frac{f_Vf_P}{M} = \left( 1 + \frac{8}{3}
\frac{\alpha_s(M)}{4 \pi} +
O(1/M) \right),
\label{eq:um}\eeq
where we take the heavy mass scale, $M$, to be the spin-averaged meson
mass, $M = (M_P+3M_V)/4$.

\begin{figure}[t]
\begin{center}
\leavevmode
\BorderBox{2pt}{%
\InsertFigure[140 480 480 800]{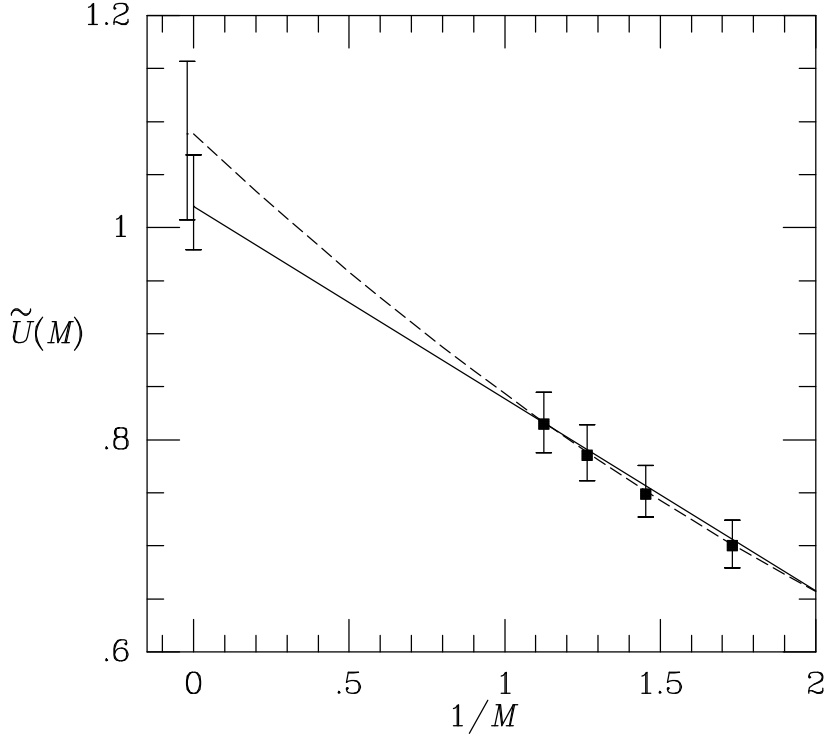}{340bp}{320bp}%
}
\begin{minipage}[t]{\captionwidth}
\caption{The quantity $\protect\tilde{U}(M)$ plotted against the inverse
spin-averaged mass. 
Linear and quadratic fits are represented by the solid and dashed
curves, respectively. Also shown are the statistical errors of the
extrapolation to the infinite mass limit.\label{fig:umext}}
\end{minipage}
\end{center}
\end{figure}

In order to test the predicted behaviour of~$U(M)$, we take the
chirally-extrapolated values for both the pseudoscalar and vector
decay constants, and fit 
\beq
\tilde{U}(M) \equiv U(M)/\left\{1 + \frac{8}{3} 
\frac{\alpha_s(M)}{4 \pi}\right\}
\eeq
to either a linear or quadratic function of $1/M$.  The data together
with the fits are shown in Fig.~\ref{fig:umext}, and we display our
results in Table~\ref{tab:uinfty}.  The perturbative values of $Z_A$
and $Z_V$ are used.
%
%

\begin{table}
\centering
\begin{tabular}{|c||c|c|} \hline
 $M$ & {Linear Fit} & {Quadratic Fit} \\ 
\hline
$\infty$           & 1.02\er{5}{4} & 1.09\er{7}{8}  \\ 
\hline
$(M_B+3\,M_B^*)/4$ & 0.93\er{4}{3} & 0.96\er{4}{5}  \\
\hline
$(M_D+3\,M_D^*)/4$ & 0.77\er{2}{2} & 0.77\er{2}{2}  \\
\hline
\end{tabular}
\begin{center}
\begin{minipage}[t]{\captionwidth}
\caption{The quantity $\tilde{U}(M)$ obtained from linear and quadratic fits.
\label{tab:uinfty}
}
\end{minipage}
\end{center}
\end{table}

The fact that $\tilde{U}(\infty )$ is around one in
Table~\ref{tab:uinfty} provides support for our parametrisations, in
eqs.~(\ref{eq:linfit}) and (\ref{eq:quadfit}), of
the non-scaling behaviour of the decay constants for finite
heavy-quark masses.


\section{Decay Constants from the Simulation at $\beta =6.0$}
\label{sec:cmclover}

In this section we describe the results of a computation of the decay
constants using the SW fermion action at $\beta=6.0$ on a
$16^3\times 48$ lattice. These results were obtained using 36
configurations, with light-quark masses corresponding to $\kappa_l =
0.1432, 0.1440$ and 0.1445. The corresponding light-light pseudoscalar
and vector meson masses, and pseudoscalar decay constants, all in
lattice units, are presented in Table \ref{tab:cmlight}. The values of
the hopping parameter corresponding to the chiral limit and the
strange quark mass
%
\begin{table}
\centering
\begin{tabular}{|c|c|c|c|} \hline
$\kappa_l$ & $m_\pi$ & $m_\rho$ & $f_\pi/Z_A$ \\ \hline 
0.1432 & 0.386\er{4}{4} & 0.51\er{2}{1} & 0.088\er{2}{3} \\ 
0.1440 & 0.311\er{6}{5} & 0.47\er{3}{2} & 0.080\er{2}{4} \\ 
0.1445 & 0.257\er{5}{6} & 0.43\er{6}{3} & 0.075\er{2}{5} \\ \hline 
$\kcrit$ = 0.14556\er{6}{6} & --- & 0.38\er{5}{4} & 0.065\er{2}{6}
\\ \hline
\end{tabular}
\begin{center}
\begin{minipage}[t]{\captionwidth}
\caption{Masses of light-light pseudoscalar and vector mesons,
and the pseudoscalar decay constants at $\beta=6.0$.
\label{tab:cmlight}
}
\end{minipage}
\end{center}
\end{table}
%
are $\kcrit=0.14556\er{6}{6}$ and $\kstrange=0.1437\er{4}{5}$
respectively. Using the mass of the $\rho$ meson to determine the
value of the lattice spacing, we find $a^{-1}=2.0\er{3}{2}\,\gev$,
whilst using $f_\pi$ we find $a^{-1}=2.1\er{2}{1}\,\gev$. These two
results are compatible, and below we will use the value
\beq
a^{-1} = 2.0\er{3}{2}\,\gev
\label{eq:scale60}\eeq
to convert the results from lattice to physical units.
\par We have computed the heavy-light correlation functions as series in 
$\kappa_h$ (the hopping-parameter expansion~\cite{hopping}), thus
enabling us to obtain the decay constants at any value of the mass of
the heavy quark, without explicitly computing the heavy-quark
propagators.  The decay constants are obtained by fitting to
eq.~(\ref{eq:psps}) and eq.~(\ref{eq:apoverpp}), over the range $12
\leq t \leq 18$ for both fits.  We employ the Jacobi smearing
algorithm with $N = 50$, corresponding to a smearing radius of $r=4.2$.

In an attempt to improve our understanding of the discretisation
errors, we have also computed the decay constants for the Wilson
action at one value of the light-quark mass, using a subset of 16 of the
36 configurations. The comparison of the results for the two actions
is presented in Subsection~\ref{subsec:cmcomp}.

\subsection{Pseudoscalar Decay Constants}
\label{subsec:cmfd}
\begin{figure}[t]
\begin{center}
\leavevmode
\BorderBox{2pt}{%
\InsertFigure[140 480 480 800]{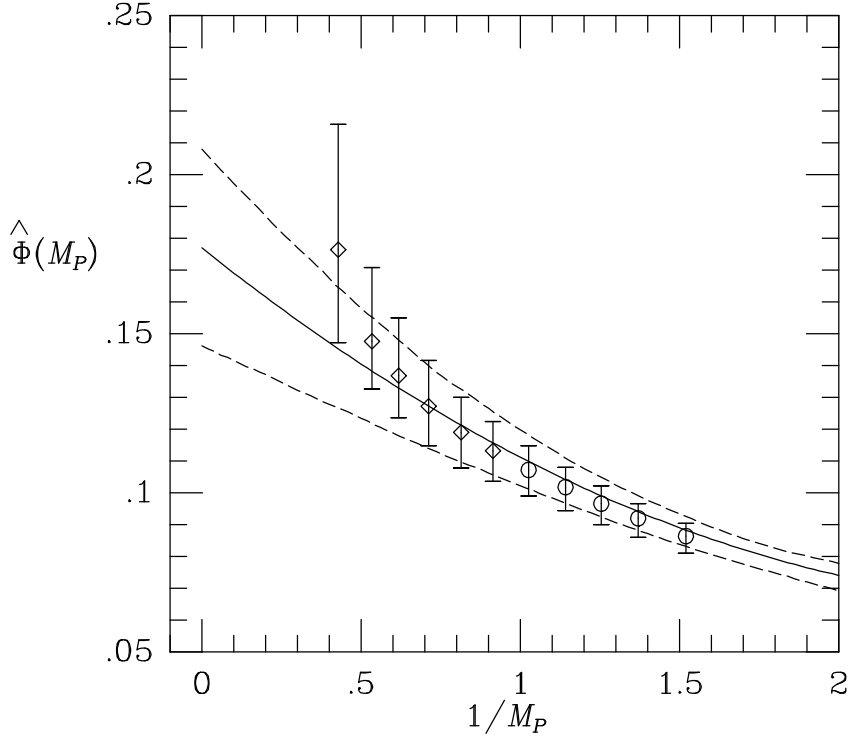}{340bp}{320bp}%
}
\begin{minipage}[t]{\captionwidth}
\caption{The chirally-extrapolated data for $\protect\hat{\Phi}(M_P)$ at $\beta=6.0$ 
plotted against the inverse meson mass. The solid curve represents a
quadratic fit to the points denoted by circles. Points represented by
diamonds are not included in the fit. The dashed curves are the 68\%
confidence bounds on the fit.\label{fig:cmfdsqrtm}}
\end{minipage}
\end{center}
\end{figure}

In Fig.~\ref{fig:cmfdsqrtm} we plot the chirally-extrapolated values of
$\hat{\Phi}(M_P)$ as a function of
$1/M_P$, for 11 values of the heavy-quark mass.  We fit the points
corresponding to the five lightest meson masses (for which
$m_Q=1/2(1/\kappa_h - 1/\kcrit)< 0.7$, as was the case at $\beta =
6.2$) to eq.~(\ref{eq:quadfit}), and this is shown as the solid curve in
the figure. For the coefficients of the fit we find:
\beq
C = 0.18\er{3}{3}\ ; \qquad D=0.45\err{13}{5}\ ;
\qquad E=0.08\er{4}{9}.\label{eq:quadfit6}\eeq
In physical units we obtain
\beqn
Ca^{-3/2} & = & 0.50\err{~9}{~9}\err{12}{~7}\, \, \, \gev^{3/2}
\nonumber \\
Da^{-1} & = & 0.91\err{21}{26}\err{14}{~9} \, \, \, \gev \nonumber \\
Ea^{-2} & = & 0.32\err{16}{36}\err{10}{~6} \, \, \, \gev^2
\label{eq:physical_fit_params60}
\eeqn
in good agreement with the results at $\beta = 6.2$, quoted in
eq.~(\ref{eq:physical_fit_params62}).  It should be noted that the
inclusion of all 11 data points makes no significant difference to
the fit.

The values for the decay constants in physical units are:
\beqn
f_D & = & 199\err{14}{15}\err{27}{19}\,\,\mev\label{eq:cmfd}\\ 
f_B & = & 176\err{25}{24}\err{33}{15}\,\,\mev\label{eq:cmfb}\\ 
\frac{f_{D_s}}{f_D} & = & 1.13\er{6}{7}\label{eq:cmfdfds}\\ 
\frac{f_{B_s}}{f_B} & = & 1.17\err{12}{12}\label{eq:cmfbfbs}\;.
\eeqn
All these numbers are in good agreement with the corresponding results
from the simulation at $\beta=6.2$. Finally, for the ratios $f_D/f_\pi$
and $f_B/f_\pi$ we obtain
\beqn
\Big(\frac{f_D}{f_\pi}\Big) \times 132\,\,\mev & = & 211\err{26}{11}\,\,\mev \\ 
\Big(\frac{f_B}{f_\pi}\Big) \times 132\,\,\mev & = & 186\err{35}{21}\,\,\mev.
\eeqn

%
%


\subsection{Comparison of Results Using Wilson and SW Actions}
\label{subsec:cmcomp}
For 16 of the configurations, we have computed the decay constants for
the Wilson fermion action, again using the hopping-parameter
expansion.  We compute light-quark propagators at a single value of
the hopping parameter, $\kappa_l^W = 0.155$, corresponding to a
pseudoscalar-meson mass of $0.30\er{1}{1}$.  This was chosen to match
the SW pseudoscalar-meson mass of $0.31\er{2}{1}$ obtained at
$\kappa_l^{\rm SW} = 0.144$ on the same set of configurations.

%
\begin{figure}[t]
\begin{center}
\leavevmode
\BorderBox{2pt}{%
\InsertFigure[140 480 480 800]{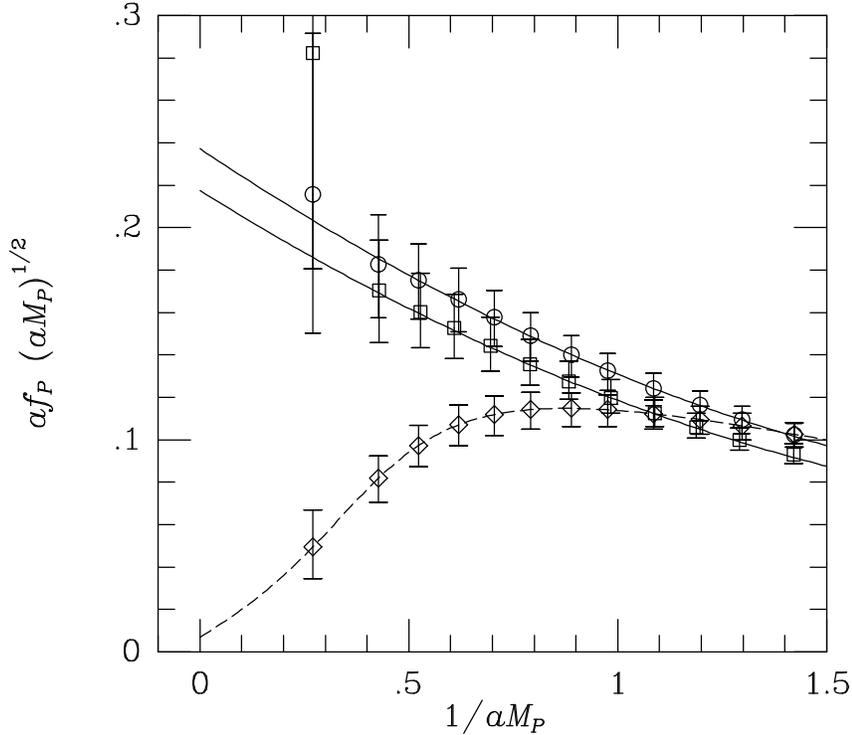}{340bp}{320bp}%
}
\begin{minipage}[t]{\captionwidth}
\caption{$f_P\protect\sqrt{M_P}$ for both the
Wilson and SW actions. Diamonds denote points obtained with the Wilson
action in the conventional normalisation, $\protect\sqrt{2\kappa}$,
whereas squares denote points normalised by
$\protect\sqrt{1-6\tilde\kappa}$.  Results using the SW action are
represented by circles.  The solid curves are quadratic fits in
$1/M_P$ to $f_P\protect\sqrt{M_P}$ for the Wilson action, with fields
normalized by $\protect\sqrt{1-6\tilde\kappa}$, and for the SW action.
The dashed curve is to guide the eye.
\label{fig:fsqrt60}}
\end{minipage}
\end{center}
\end{figure}
%

\par In Fig.~\ref{fig:fsqrt60} we plot  $f_P \sqrt{M_P}$ as
a function of $1/M_P$.  Using the conventional normalisation of
$\sqrt{2\kappa}$ for the quark fields, we see a clear divergence
between the results for the two actions for $m_Q > 0.7$; the Wilson
results turn over and decrease.
However, we note that uncorrelated $\chi^2$
fits of the Wilson points,
at the lightest few meson masses, to eqs.~(\ref{eq:linfit}) and
(\ref{eq:quadfit})
would yield coefficients of the $1/M_P$
term broadly consistent with previous Wilson analyses.
\par The figure also shows the results obtained
with the Wilson action, but using the normalisation
$\sqrt{1-6\tilde\kappa}$ for the quark fields
\cite{labrenz,lm,luscher}, where $\tilde\kappa=u_0\kappa$ and
$u_0=1/(8\kcrit)$. It has been suggested that this normalisation may
absorb some of the discretisation errors \cite{lm}, and indeed the
corresponding results agree remarkably with those obtained using the
SW action.  This agreement provides considerable motivation for a
theoretical study to investigate whether there is any formal
connection between the ansatz above and the improvement programme
initiated by Symanzik~\cite{symanzik}.

\section{$f_B$ in the Static Limit}
\label{sec:static}

%
An alternative and complementary approach to heavy-quark physics 
using lattice QCD was proposed by Eichten \cite{eichten}. This technique
is based on an expansion of the heavy-quark propagator in inverse powers
of the quark mass. In practice, one keeps just the leading
term, given by (at zero velocity)
\beq
S_Q(\vec x,t;\vec 0, 0)=\left\{\theta(t)e^{-m_Qt}\frac{1+\gamma^4}{2} + 
\theta(-t)e^{m_Qt}\frac{1-\gamma^4}{2}\right\}\,
\delta^{(3)}(\vec x)\calp _{\vec 0}(t,0),
\label{eq:sb0}\eeq
where $\calp _{\vec 0}(t,0)$ is the product of links from $(\vec 0,t)$ 
to the origin, for example for $t>0$,
\beq
\calp _{\vec 0}(t,0) = U_4^\dagger(\vec 0,t-1)
U_4^\dagger(\vec 0, t - 2) \cdots U_4^\dagger(\vec 0, 0).
\label{eq:calp}\eeq 
At sufficiently large times 
\beq
\sum_{\vec x} \langle A_4(\vec x, t)A_4^\dagger(0)\rangle \to \frac 
{f_P^2M_P}{2} e^{-M_Pt},
\label{eq:fbstatic}\eeq
where $A_\mu$ is the improved axial current of
eq.~(\ref{eq:static_rotations}) with $\Gamma = \gamma_\mu \gamma_5$.
Since the only dependence on $m_Q$ in eq.~(\ref{eq:fbstatic}) arises
through the exponential factor in eq.~(\ref{eq:sb0}), we deduce the
scaling law that $f_P\,\sqrt{M_P}$ is independent of the heavy-quark
mass.  Matching the result from the Heavy Quark Effective Theory with
that in the full theory introduces the logarithmic corrections in
eq.~(\ref{eq:fP_scaling}).  The full scaling law is of the form
\beq
f_P \sqrt{M_P} = {\rm const.} \left[ (\alpha_s(M_P))^{-2/\beta_0}(1+O(\alpha_s)) 
+\, O(1/M_P) \right].
\label{eq:fpstatic}\eeq
The objective of lattice computations is to determine the constant.
We refer to the value of $f_B$ obtained using eq.~(\ref{eq:fpstatic}),
but dropping the $O(1/M_B)$ corrections, as $\fbstat$.
\par We compute the two correlation functions, $C^{SS}$ and $C^{LS}$, defined
by\footnote{In principle the behaviour of the correlation functions in
eqs.~(\ref{eq:css}) and (\ref{eq:cls}) is given by a cosh (as in 
eq.~(\ref{eq:psps})), however the contribution of the backward-propagating
meson is negligible in the time intervals we will be considering.}
\beqn
C^{SS}(t) & = & \sum_{\vec x} \langle0| A_4^S(\vec x,t) A_4^{\dagger S}
(\vec 0,0)|0\rangle\rightarrow (Z^S)^2 e^{-\Delta E\,t}\label{eq:css}\\ 
C^{LS}(t) & = & \sum_{\vec x} \langle0| A_4^L(\vec x,t) A_4^{\dagger S}
(\vec 0,0)|0\rangle\rightarrow Z^L Z^S e^{-\Delta E\,t}\label{eq:cls},\eeqn
where $\Delta E$ is the (unphysical) difference between the mass of
the meson and the bare mass of the heavy quark.  The matrix element of
the local operator $A_4^L$ is obtained from the two correlation
functions $C^{SS}$ and $C^{LS}$ as follows.  By fitting $C^{SS}(t)$ to
the functional form given in eq.~(\ref{eq:css}) we obtain $Z_S$ (and
$\Delta E$).  At sufficiently large times the ratio
$C^{LS}(t)/C^{SS}(t)\rightarrow Z^L/Z^S$, so that $Z^L$ can be
determined.

\subsection{Results at $\beta = 6.2$}\label{subsec:static62}
\par We now report on a computation of $\fbstat$ at $\beta = 6.2$.
The results presented here were obtained using a subset of 20
of the 60 configurations discussed in Section \ref{sec:fdprop},
at the three values of the light-quark mass.
The values of $\fbstat$ and $\fbsstat$
were determined by extrapolating the results to $\kcrit$ and
$\kstrange$ respectively.
\par In view of the difficulty in isolating the ground state in correlation 
functions using the static effective theory, we have compared results
obtained with different numbers of iterations of the Jacobi smearing
algorithm~\cite{bew}.  For $N$ less than about 80 the plateaus do not
start until at least $t=7$.  In this paper we present our results
obtained with $N=110$ and $N=140$, corresponding to $r=5.9$ and 6.4
respectively, where plateaus begin as early as $t=4$ and hence
statistical errors are smaller.
\par
\begin{figure}[t]
\begin{center}
\leavevmode
\BorderBox{2pt}{%
\InsertFigure[110 250 490 570]{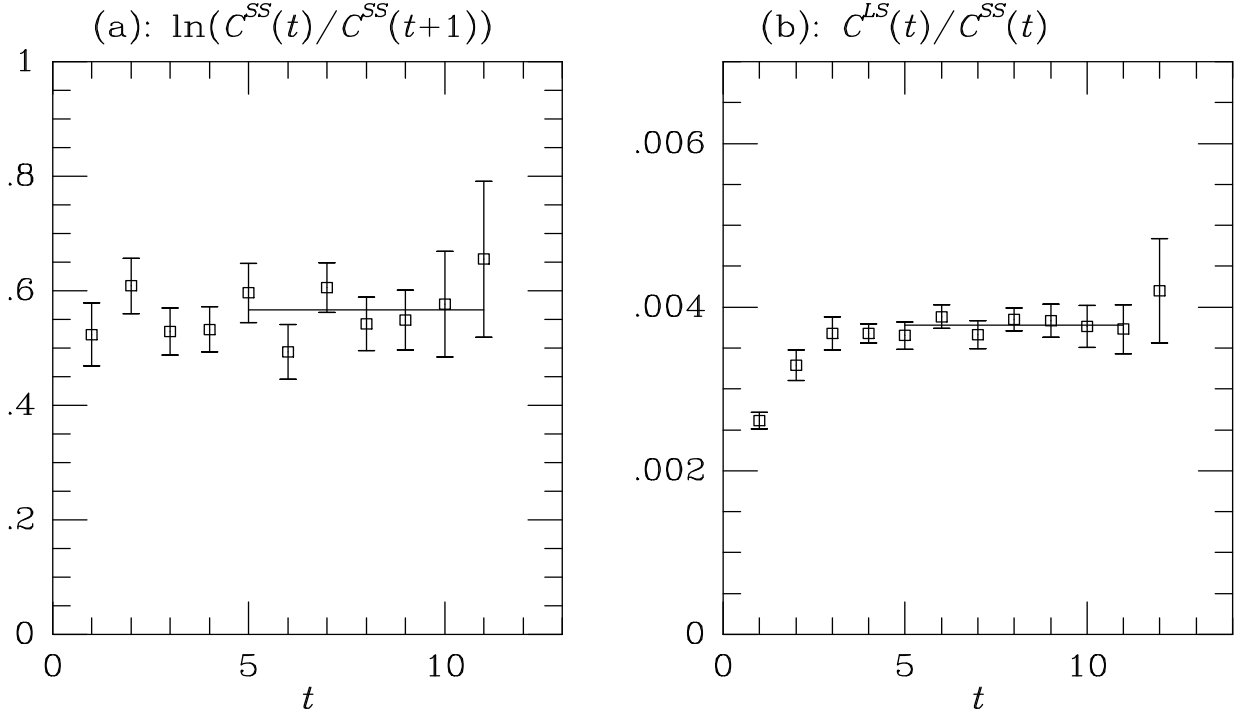}{400bp}{320bp}%
}
\begin{minipage}[t]{\captionwidth}
\caption{(a) The effective mass obtained from $C^{SS}(t)$, and
(b) the ratio $C^{LS}(t)/C^{SS}(t)$ at $\beta=6.2$,
$\kappa_l=0.14226$ and $N = 140$. The
solid lines represent fits from $5\leq t \leq11$.
\label{fig:fbstat}}
\end{minipage}
\end{center}
\end{figure}

 In Fig.~\ref{fig:fbstat}(a) we show 
the effective masses obtained from $C^{SS}(t)$,
and in Fig.~\ref{fig:fbstat}(b)
the ratio $C^{LS}(t)/C^{SS}(t)$, both at $\kappa_l = 0.14226$ and $N = 140$.
Excellent
plateaus are obtained, giving us confidence that the ground state has
indeed been isolated.
%
%
In Table \ref{tab:fbstat} we present the  results for $\Delta E$, $(Z^S)^2$, 
$Z^L/Z^S$ and $Z^L$ at all three values of $\kappa_l$, from fits over
the range $5 \leq t \leq11$, without symmetrization in Euclidean time.
$\Delta E$ is obtained from the fit to $C^{SS}(t)$ for $N = 140$;
consistent values are obtained for $N = 110$.
%
\begin{table}
\centering
\begin{tabular} {|c||c||c|c||c|c||c|c|}\hline
$\kappa_l$ & $\Delta E$ & \multicolumn{2}{c||}{$(Z^S)^2$} &
\multicolumn{2}{c||}{$Z^L/Z^S$} & \multicolumn{2}{c|}{$Z^L$} \\ \hline
& $N=140$ & $N=110$ & $N=140$ & $N=110$ & $N=140$ & $N=110$ & $N=140$ \\
\hline 0.14144 & 0.59 \er{1}{1} & 141\err{12}{11} & 130\err{10}{11} & 0.0125\er{3}{3} &
0.0039\er{1}{1}& 0.149\er{7}{7} & 0.142\er{7}{6} \\ 
0.14226 & 0.57\er{1}{1} &
127\err{12}{11} & 119\err{10}{11} & 0.0121\er{3}{3} & 0.0038\er{1}{1}&
0.137\er{7}{6} & 0.130\er{7}{6} \\ 
0.14262 & 0.56\er{2}{1} &
119\er{12}{11} & 112\err{9}{11} & 0.0120\er{3}{3} & 0.0037\er{1}{1}&
0.131\er{7}{6} & 0.125\er{7}{7} \\
\hline
\end{tabular}
\begin{center}
\begin{minipage}[t]{\captionwidth}
\caption{Values of $\Delta E$, $(Z^S)^2$, $Z^L/Z^S$ and $Z^L$ at the
three value of $\kappa_l$.  $\Delta E$ is obtained from the fit to
$C^{SS}(t)$.
\label{tab:fbstat}
}
\end{minipage}
\end{center}
\end{table}
%
\par Extrapolating the results for $Z^L$ in Table \ref{tab:fbstat} to the 
chiral limit and to the mass of the strange quark we find:
\beqn
 Z^L & = & 0.124\er{8}{7}\ \ \ \ {\rm at}\ \ \kappa_l=\kcrit 
\label{eq:zlc110}\\ 
 Z^L & = & 0.140\er{7}{6}\ \ \ \ {\rm at}\ \ \kappa_l=\kstrange  
\label{eq:zls110}\eeqn
when obtained using smeared interpolating operators with $N=110$ and
\beqn
 Z^L & = & 0.117\er{7}{7}\ \ \ \ {\rm at}\ \ \kappa_l=\kcrit 
\label{eq:zlc140}\\ 
 Z^L & = & 0.134\er{7}{6}\ \ \ \ {\rm at}\ \ \kappa_l=\kstrange  
\label{eq:zls140}\eeqn
when using interpolating operators with $N=140$. 

When matching the static lattice theory to the full theory at a scale
$m_b$, the factor required is~\cite{eichill}:
\beq
Z_A^{{\rm stat}} = {\cal Z}_A
\left(1+\frac{\alpha_s(a^{-1})}{3\pi}
\Big[\frac{3}{2}\log a^2m_b^2 -2 \Big]
\right).\label{eq:zstatfull}
\eeq
${\cal Z}_A$, relating the axial 
current in the static lattice theory to the static continuum one, has
been calculated in perturbation theory for the SW
action~\cite{hernandez,borrelli2}:
\beq
{\cal Z}_A = 1 - 0.127\, g^2 
\simeq 0.79.
\label{eq:zastat}\eeq
The value of 0.79 on the right hand side of eq.~(\ref{eq:zastat}) was
estimated using the boosted coupling at $\beta=6.2$.  For the
remaining factor in eq.~(\ref{eq:zstatfull}), we take $m_b=5\,\gev$,
$\alpha_s$ given by eq.~(\ref{eq:alphas}) with $n_f = 0$, and
$\Lambda_{\rm QCD} = 200\,
\mev$, yielding a number close to one (note that this is insensitive to
small changes in $m_b$).  Thus $Z_A^{{\rm stat}} = 0.79$ also.  We
find
\beqn
\fbstat  & = & 266 \err{17}{15}\errr{110}{14}
\left(\frac{Z_A^{{\rm stat}}}{0.79}\right)\,\mev\\ 
\fbsstat & = & 300 \err{14}{13}\errr{125}{16}
\left(\frac{Z_A^{{\rm stat}}}{0.79}\right)\,\mev,
\eeqn
when using the interpolating operators with $N=110$, and 
\beqn
\fbstat  & = & 253 \err{16}{15}\errr{105}{14}
\left(\frac{Z_A^{{\rm stat}}}{0.79}\right)\,\mev\label{eq:fbstatres}\\ 
\fbsstat & = & 287 \err{14}{13}\errr{119}{15}
\left(\frac{Z_A^{{\rm stat}}}{0.79}\right)\,\mev
\label{eq:fbsstatres}\eeqn
when using those with $N=140$. The systematic errors quoted arise from
the uncertainty in the scale. We take the results in equations 
(\ref{eq:fbstatres}) and (\ref{eq:fbsstatres}) as our best values, and
these give for the ratio:
\beq
\frac{\fbsstat}{\fbstat} = 1.14\er{4}{3}.\label{eq:fbstatrat62}
\eeq

\subsection{Results at $\beta = 6.0$}\label{subsec:static60}
We have performed a similar analysis on the 36 configurations at
$\beta = 6.0$, using Jacobi smearing with $N = 50$ and 150,
corresponding to $r = 4.2$ and 6.2 respectively.  The results obtained
using the two smearing radii are consistent, and our best results are
those at $N = 50$ for which $Z_L = 0.211\er{6}{7}$, yielding
\beqn
\fbstat & = & 286 \err{~8}{10}\err{67}{42}
\left(\frac{Z_A^{{\rm stat}}}{0.78}\right)\,\mev\label{eq:fbstatres6}\\
\fbsstat & = & 323 \err{14}{14}\err{75}{47}
\left(\frac{Z_A^{{\rm
stat}}}{0.78}\right)\,\mev.\label{eq:fbsstatres6} \\
\frac{\fbsstat}{\fbstat} & = & 1.13\er{4}{3}.\label{eq:fbstatrat60}
\eeqn
These results at $\beta = 6.0$ are consistent with those at $\beta =
6.2$ presented in eqs.~(\ref{eq:fbstatres}), (\ref{eq:fbsstatres}) and
(\ref{eq:fbstatrat62}).

%
\begin{figure}[t]
\begin{center}
\leavevmode
\BorderBox{2pt}{%
\InsertFigure[110 250 490 570]{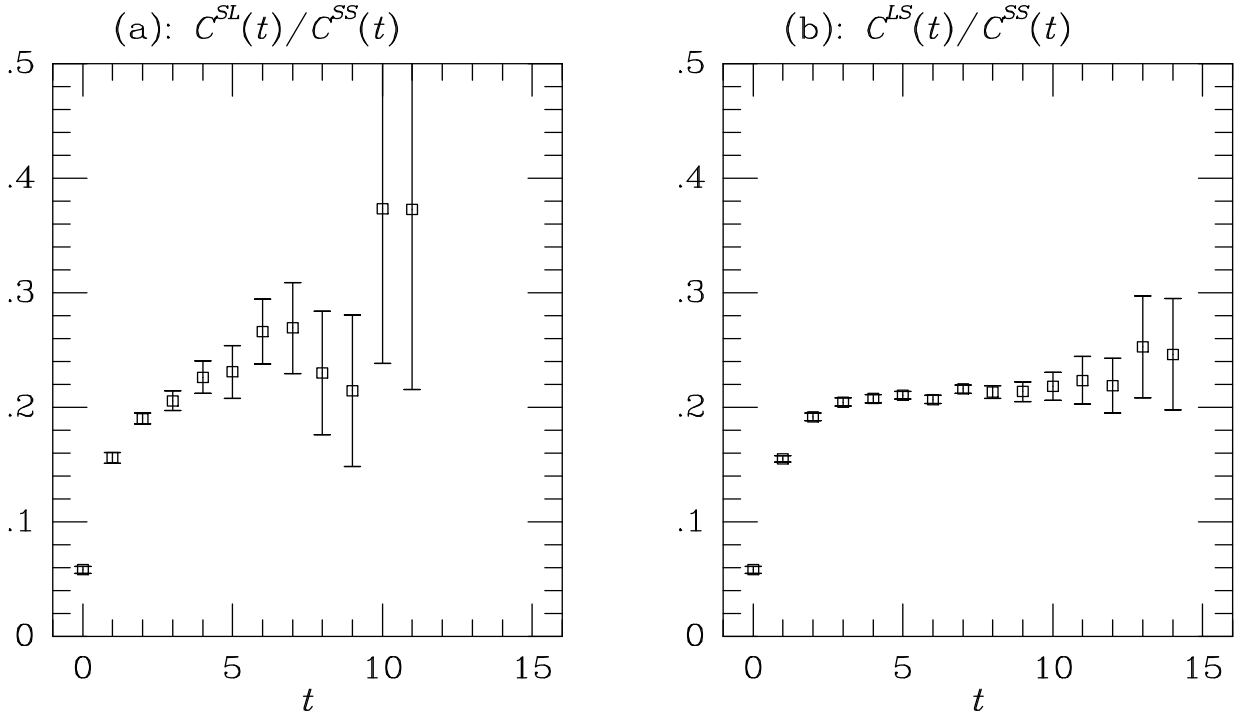}{400bp}{320bp}%
}
\begin{minipage}[t]{\captionwidth}
\caption{(a) The ratio $C^{SL}(t)/C^{SS}(t)$ and (b)
$C^{LS}(t)/C^{SS}(t)$ plotted against~$t$ for $\beta=6.0$,
$\kappa_l=0.1432$, and $N=50$ iterations used in the Jacobi smearing
algorithm.
\label{fig:fbstat60}}
\end{minipage}
\end{center}
\end{figure}
%
\par The results plotted in Fig.~\ref{fig:fbstat}(b) for the ratio 
$C^{LS}(t)/C^{SS}(t)$ appear to have considerably smaller errors, and
a clearer plateau, than in some recent
studies~\cite{allton,apesmearing}, in spite of our limited statistics.
We attribute this to the fact that we use the $C^{LS}$ correlation
function in which the smearing is performed at the source, rather than
the $C^{SL}$ correlation function in which the smearing is performed
at the sink\footnote{In both refs.~\cite{allton} and
\cite{apesmearing} it was in fact the $C^{SL}$, and not the $C^{LS}$,
 correlation function
which was computed~\cite{guido}.}.  Of course, with sufficiently many
configurations, the results are independent of this choice. However in
the $C^{LS}$ correlation function, the heavy-quark propagator is
sampled at many spatial points, whereas in the $C^{SL}$ correlation
function only the heavy-quark propagator at $\vec x = \vec 0$
contributes. Thus it seems plausible that the statistical errors are
considerably reduced using the $C^{LS}$ correlation function.
\par To
check this hypothesis, we have computed the ratios
$C^{LS}(t)/C^{SS}(t)$ and $C^{SL}(t)/C^{SS}(t)$
at $\beta = 6.0$, with $\kappa_l=0.1432$. The results
are plotted in Fig.~\ref{fig:fbstat60}, and indeed confirm that there
is an enormous improvement in precision when the correlation function
$C^{LS}$ is used. We believe that this, rather than the different
method of smearing, is the reason for the relatively poor plateaus in
ref.~\cite{apesmearing}.

\subsection{Discussion}\label{subsec:statdiscussion}
\par
\begin{figure}[bht]
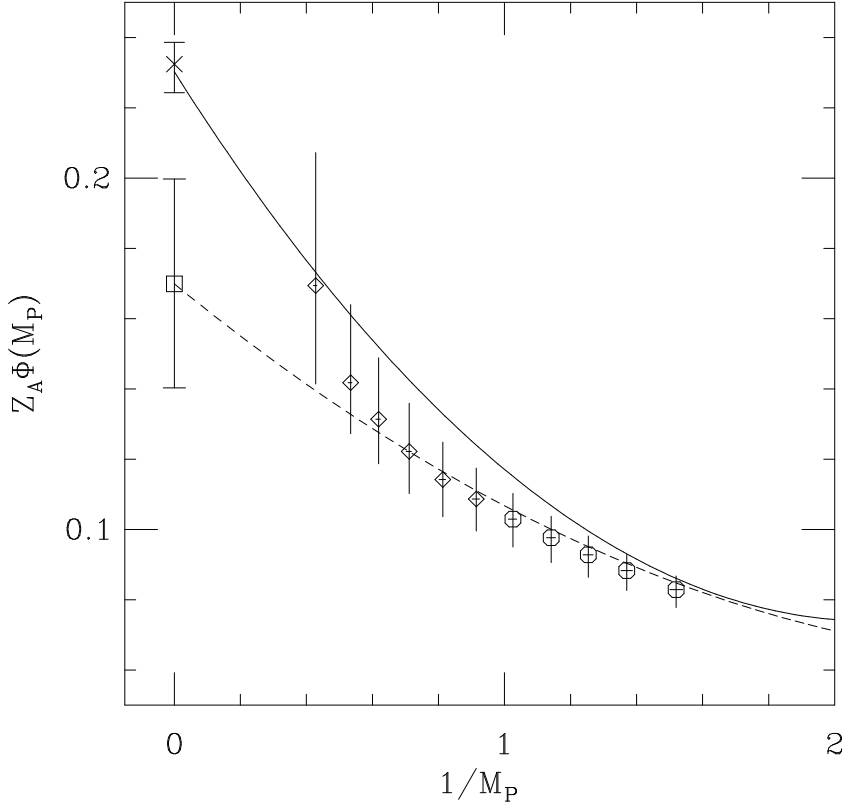

\begin{center}
\leavevmode
\BorderBox{2pt}{%
\InsertFigure[20 30 620 600]{fball.ps}{340bp}{320bp}%
}
\begin{minipage}[t]{\captionwidth}
\caption{$Z_A\protect\hat{\Phi}(M_P)$ at $\beta = 6.0$
from the simulation with propagating quarks (open symbols) and the
static theory (cross).  The dashed curve is the fit to the open
circles, with the parameters of eq.~(\protect\ref{eq:quadfit6}); the
square is the intercept at $1/M_P = 0$. The
solid curve is the fit with the static point included.
\label{fig:fball}}
\end{minipage}
\end{center}
\end{figure}

 We begin with a comparison of the static and propagating results.
Because we do not yet have static results for the full set of
configurations at $\beta = 6.2$, we focus on a comparison at $\beta =
6.0$.  In Fig.~\ref{fig:fball} we plot our results for the scaling
quantity $f_P\sqrt{M_P} (\alpha(M_P)/\alpha(M_B))^{6/33}$ from the
simulation at $\beta=6.0$ as a function of $1/M_P$ (in lattice units),
together with our result for $\fbstat\sqrt{M_B}$.  The quadratic fit
which we used to obtain our estimate for $f_B$ in
Subsection~\ref{subsec:cmfd} gives an intercept at $1/M_P=0$ which is
about 25\% and two standard deviations below the static result; a
similar discrepancy is observed at $\beta = 6.2$. There are a number
of possible reasons for this, e.g.  uncertainties in the
renormalisation constants (which are different for the static and
propagating quarks), residual discretisation errors in the simulation
of the propagating quarks, and uncertainties in the various
extrapolations.  

\par A better way of determining the consistency of the static and
propagating results is to include the static result in the
quadratic fit.  Such a fit using the full correlation matrix at $\beta
= 6.0$ gives a $\chi^2/{\rm dof}$ of 1.5.  This is still
acceptable, and provides further evidence that using rotated
operators with the SW action gives a sensible normalisation for
propagating heavy-quark fields.  From this fit we obtain $f_B =
220\er{6}{7}\err{40}{27}$ \mev, which is 44 $\mev$ higher than that
obtained from the propagating points alone.

\par In Table \ref{tab:fbstatcomp} we present the results for $\fbstat$
obtained by other groups, together with our values. Although at $\beta
= 6.0$ the values of $Z^L$ found by all groups and for both actions
are in broad agreement, the different treatment of systematics leads
to the spread of results in $\fbstat$.
\begin{table}
\centering
\begin{tabular} {|c|c|c|c|c|l|} \hline
Ref. & Action & $\beta$ & $a^{-1}$ [\gev] & $Z_A^{{\rm stat}}$ 
& \multicolumn{1}{|c|}{$\fbstat$ [\mev]}\\ 
\hline
\cite{duncan} & Wilson & 5.9 & 1.75 & 0.79 & $319\pm 11$ \\ 
\cite{allton} & Wilson & 6.0 & $2.0\pm 0.2$ & 0.8 & $310\pm 25\pm 50$ \\ 
\cite{wuppfbstat} & Wilson & 6.0 & $2.2\pm 0.2$  & 0.8 & $366\pm 22\pm 55$ \\ 
\cite{apesmearing} & Wilson & 6.0 & $2.11\pm 0.05\pm 0.10$  & 0.8 &
$350 \pm 40 \pm 30$ \\ 
\cite{apesmearing} & SW & 6.0 & $2.05\pm 0.06$  & 0.89 &
$370 \pm 40$ \\ 
This Work & SW & 6.0 & $2.0\er{3}{2}$  & 0.78 & 
$286\err{~8}{10}\err{67}{42}$\\ 
This Work & SW & 6.2 & $2.7\er{7}{1}$  & 0.79 & 
$253\err{16}{15}\errr{105}{~14}$\\ 
\cite{labrenz} & Wilson & 6.3 & $3.21\pm .09\pm .17$ & 0.69 & 
$235\pm 20\pm 21$ \\
\cite{alexandrou2} & Wilson & 5.74, 6.0, 6.26 & 
1.12, 1.88, 2.78  & 0.71(8) & $230\pm 22\pm 26$ \\ 
\hline
\end{tabular}
\begin{center}
\begin{minipage}[t]{\captionwidth}
\caption{Compilation of lattice results for $\fbstat$
\label{tab:fbstatcomp}
}
\end{minipage}
\end{center}
\end{table}
It has been suggested that $\fbstat$ decreases as $a\rightarrow 0$ 
\cite{alexandrou2}.  However, the agreement of the results obtained with
the Wilson and SW actions at $\beta = 6.0$, together with consistency
between our results at $\beta = 6.0 $ and $\beta = 6.2$, suggests that
the discretisation errors are small.

%
%
%
%
%
%


\section{Conclusions}
\label{sec:concs}
In this paper we have carried out an extensive study of the decay
constants of heavy-light mesons using the SW action for the quarks.
The use of the SW action confirms the large, negative $O(1/M_P)$
corrections to the scaling law $f_P\sqrt{M_P}\sim$~constant at the
mass of the charm quark and the significant corrections at the mass of
the $b$ quark, previously observed with the Wilson action. However,
from our comparison of results for the Wilson and SW actions at $\beta
= 6.0$, we observe clear evidence that the $\sqrt{2\kappa}$
normalisation of the Wilson quark fields fails for large quark mass.
This failure is presumably due to large $O(m_Q a)$ effects.  It has
been suggested that such effects may largely be absorbed by the use of
a different normalization~\cite{labrenz}.  We find that such a
normalization yields results in agreement with those obtained using
the SW action with rotated operators and the $\sqrt{2\kappa}$
normalization for the quark fields.

Our best estimates of $f_D$ and $f_B$ are
\beqn
f_D & = & 185\er{4}{3}\,({\rm stat})\,\err{42}{7}\,({\rm syst})\,\,\mev  
\label{eq:bestfdagain}\\ 
f_B & = & 160\er{6}{6}\,\err{53}{19}\,\,\mev,
\label{eq:bestfbagain}
\eeqn
obtained using propagating quarks at $\beta = 6.2$.  Our analysis at
$\beta = 6.0$ yields entirely consistent results.  The latter analysis
also suggests that including the static result in the fits is likely
to increase the value of $f_B$ by around $40\;\mev$.

The most urgent extension of this work is to determine the
$B$ parameter for $B^0$--$\bar B^0$ mixing, since it is the combination
$f_B\sqrt{B_B}$ which is directly relevant for phenomenological
studies of the mixing and of $CP$-violation. A recent simulation with
Wilson fermions found $B_B=1.16\pm 0.07$~\cite{abada}, and it is
important to confirm this result with the improved action.

\paragraph{Acknowledgements}

We are grateful to G.\,Martinelli for many helpful discussions.  CTS
(Senior Fellow) and ADS (Personal Fellow) acknowledge the support of
the Science and Engineering Research Council.  This work was supported
by SERC Grants GR/32779, GR/H 49191, GR/H 01069, GR/H 53624, the
University of Edinburgh and Meiko Limited.  We are grateful to Mike
Brown of Edinburgh University Computing Service, and to Arthur Trew of
EPCC, for provision and maintenance of service on the Meiko i860
Computing Surface and the Thinking Machines CM-200.  We wish to thank
Brian Murdoch for access to a Digital 7640 AXP (``Alpha'') system
placed at the University of Edinburgh Computing Service by Digital
Equipment Corporation for field test.




\end{document}